\documentclass[preprint,prb,aps,showkeys,amsmath,amssymb]{revtex4-1}
\usepackage{epsfig}
\usepackage{amsmath}

\usepackage{epsfig}
\usepackage{amsmath}
\usepackage{color}

\usepackage{color}
\usepackage[latin9]{inputenc}
\usepackage{float}
\usepackage{natbib}
\usepackage{graphicx}

\begin{document}

\title[Three phase equilibria of the methane hydrate in NaCl solutions: a simulation study]{Three phase equilibria of the methane hydrate in NaCl solutions: a simulation study }
\author{S. Blazquez}
 
\author{C. Vega}%
\affiliation{ 
Departamento de Qu\'{\i}mica F\'{\i}sica,
Facultad de Ciencias Qu\'{\i}micas, \\ Universidad Complutense de Madrid,
28040 Madrid, Spain.
}%

\author{M. M. Conde*}
 \email{maria.mconde@upm.es}
\affiliation{%
Departamento de Ingeniería Química Industrial y del Medio Ambiente, Escuela Técnica Superior de Ingenieros Industriales, Universidad Politécnica de Madrid,
28006, Madrid, Spain.
}%

\date{\today}

\begin{abstract}

Molecular dynamics simulations have been performed to determine the three-phase coexistence temperature for a methane hydrate system in equilibrium with a NaCl solution and a methane gas phase. The direct coexistence technique is used following two approaches, one where the triple coexistence temperature for a given NaCl concentration is narrow down and another where the concentration at a given temperature is equilibrated. In both approaches the results are consistent within the error bars. All simulations were carried out at 400 bar and the range of concentrations explored extends up to a molality of 4 m. TIP4P/2005 for water molecules and a simple Lennard-Jones interaction site for methane were used to simulate the system. Positive deviations from the Lorentz-Berthelot energetic rule have been applied between methane and water (i.e., increasing the attractive interaction between water and methane). Na$^+$ and Cl$^-$ ions were described by using the Madrid-2019 scaled charge model. The role played by finite size effects in the calculation of the coexistence line was analyzed by studying a system with larger number of molecules at a given NaCl concentration. Overall, our simulations show that upon NaCl addition to the liquid water phase, a shift in the three-phase equilibrium line to lower temperatures is produced as occurs in the ice-NaCl(aq) system. The depression of the three-phase coexistence line obtained at different concentrations is in a very good agreement with the experimental results.

\end{abstract}

\maketitle

\section{\label{sec:intro}Introduction}

Gas clathrate hydrates are non-stiochemical compounds that can accommodate small molecules in their framework of solid water molecules, generally in gas phase (e.g. N$_2$, CH$_4$ or CO$_2$) \cite{Sloan_book_hydrates,PTRSL_1811_101_1,Nature_1987_325_135,mcmullan1965polyhedral,mak1965polyhedral,alavi2022clathrate}. Gas clathrate hydrates have multiple applications ranging from energy storage \cite{N_434_743_2005,JPCB_113_7558_2009,N_2003_426_353,S_2004_306_469,JPCB_112_1885_2008,JCP_130_014506_2009,AIChE_2003_49_1300}, CO$_2$ sequestration \cite{Aya-CO2,IJT_17_1_1996,Herzog-CO2,yezdimer,libro-app-gas} or environmental implications to the presence of these compounds in icy satellites \cite{JPCC_1_80_2020,https://doi.org/10.1002/2014RG000463,PCCP_2017_19_9566,D1CP02638K,PRIETOBALLESTEROS2005491,KARGEL2000226,10.1130/G22311.1}.

In recent decades, the implications on the energy storage of these systems have received special interest and the number of published works has grown exponentially. Methane hydrates are of particular interest due to the existence of large reserves on the ocean floor \cite{Sloan_book_hydrates,CG_1988_71_41,ARE_1990_15_53,GRL_34_L22303_2007,N_2007_445_303,N_2003_426_353}. In fact, the first extractions of methane gas from offshore methane hydrate deposits have begun to take place in Japan\cite{YU20195213,C9RA00755E,10.4043/25243-MS,doi:10.1021/acs.energyfuels.6b03143}. 

Methane hydrates are stable at low temperatures and moderate pressures. Precisely, these thermodynamic conditions make the possibility of environmental catastrophes already a fact, as happened in Brazil in 2020\cite{ketzer2020gas}. The release of methane from the interior of the hydrates causes strong explosions. In this way, it is crucial to control the conditions of formation and dissociation of these systems.

Extensive research has been carried out to study the properties and behavior of these compounds. Not only experimental studies, but also molecular simulations have proven to be a very valuable tool to elucidate the questions that remain open in the formation and dissociation mechanisms of gas clathrate hydrates\cite{krishnan2022controlling,JCP_2015_142_124505,JCP_2010_133_064507,conde2010can,doi:10.1021/acs.jpclett.8b01210,doi:10.1021/ja309117d,JCP_149_074502_2018,doi:10.1143/JPSJS.81SA.SA027,JPCB_118_1900_2014,JPCB_118_11797_2014,JPCC_120_3305_2016,JPCC_123_1806_2019,JPCC_1_80_2020,JPCC_122_2673_2018,SUSILO20086,doi:10.1021/jp807208z,doi:10.1021/jp4023772,KONDORI2019264,doi:10.1021/acs.jpclett.8b01210,doi:10.1021/acscentsci.8b00755,doi:10.1021/jp504852k,Walsh1095,doi:10.1063/1.5084785,BRUMBY2016242,JCP_2015_142_124505,doi:10.1021/acs.jpcc.5b05393,doi:10.1139/cjc-2015-0003,KONDORI2017754,doi:10.1021/jp102874s,doi:10.1021/acs.jpcb.8b02285} or the mechanism of nucleation \cite{doi:10.1021/jp5002012,doi:10.1021/jz3012113,doi:10.1063/1.4866143,doi:10.1021/ja309117d,doi:10.1021/jp107269q,doi:10.1021/acs.cgd.0c01303,doi:10.1021/jp507959q}. The use of good force fields to reproduce the properties of water and gas is crucial. A first step to test these models is to determine the three-phase coexistence line (hydrate-liquid-gas). The knowledge of the thermodynamic stability conditions of the system is very important to be certain that the simulated phase corresponds to the stable phase. The values of the equilibrium temperature and pressure in simulation do not always correspond to experimental values, and for this reason it is essential when one wants to simulate this type of system to know the range of stability of the model employed. Likewise, the equilibrium temperature of the three phases in hydrates is crucial for nucleation studies, as it is necessary to know the degree of subcooling when the critical size cluster begins to nucleate. Additionally, having all the variables of the three-phase equilibrium line under control is essential when implementing environmental safety measures.

In 2010, we presented one of our first works in the field of gas hydrates studying the three-phase equilibrium line of methane hydrate by computer simulations\cite{JCP_2010_133_064507,conde2013note}. The liquid phase consisted solely of pure water without electrolytes dissolved in it. However, in order to replicate the conditions of the seabed or to reproduce the conditions of the icy planetary satellites, it is necessary to take into account the presence of electrolytes, specifically sodium chloride (NaCl) due to its abundance. Methane hydrates in salty aqueous solutions are of great interest \cite{JPCB_118_11797_2014,QI20126,SEO2019980,doi:10.1021/acs.jced.8b00155,HU201827,HU2017750,doi:10.1021/acs.jced.7b00292,CHA20162,LAFOND20121,doi:10.1021/acs.jpcc.8b03154,doi:10.1021/je500841b,doi:10.1021/acs.energyfuels.5b01416,SUN201892,NGUYEN201587,doi:10.1021/acs.energyfuels.9b00490,YANG2019266,https://doi.org/10.1002/2016GL072277,SR_2019_9_5860,doi:10.1021/acs.energyfuels.8b03486,LI2020125126,https://doi.org/10.1002/aic.15846,QASIM2020116219,doi:10.1021/ef402445k,doi:10.1021/jp308224v,doi:10.1021/acs.cgd.8b01161,SUN2015116,doi:10.1080/15567036.2011.557695,doi:10.1021/acs.jpca.0c00621,JAGER200189} and recently they have been studied using machine learning techniques\cite{XU2021107358}. Thus, in this paper we focus on the study of the equilibrium line of three phases of methane hydrate where the liquid phase is a solution of sodium chloride. For a phase of pure ice in contact with a sodium chloride solution, the equilibrium temperature decreases as the concentration of salt in solution increases\cite{JML_2018_261_513,lamas2022freezing}. Freezing point depression is produced by the addition of salt. An analogous behavior is expected in the case of methane hydrates since the framework of water molecules of the hydrate is considered to be a similar type of ice phase.

More than 10 years have been necessary to present this new study of the equilibrium line of three phases in methane hydrates. Although the only new element is the addition of salt to the system, from the point of view of computational simulation it involves many more details. For example, the simulation time from one system to another is increased by an order of magnitude by the mere presence of the ions that slow down the diffusion coefficient of water and the dynamics of the process significantly. 

Ionic solutions can play a key role in simulations\cite{holovko1991effects,holovko2001influence,kalyuzhnyi1993analytical}.
Thus choosing a potential model that correctly reproduces the properties of ions in aqueous solution is not an easy task. Recently, a new family of ion models has appeared based on the use of scaled charges for the ions\cite{JCP_2019_151_134504,madrid_2019_extended,fue:jpc16,fue:pa18,koh:jpcb15,mar:jcp18,li:jcp15,kan:jcp14}.
In these models the charge of the cations (for instance Na$^+$)  is not an integer number in electron units but a fraction of it (typically $\pm$0.85\textit{e} or $\pm$0.75\textit{e}) and the same is true for the anions (for instance Cl$^-$).  
These models reveal promising results in determining the equilibrium properties of ions in water. For our present work we shall use the Madrid-2019 force field\cite{JCP_2019_151_134504,madrid_2019_extended} to describe an aqueous solution of NaCl in water. This force field uses the concept of scaled charges (with a choice of $\pm$0.85\textit{e}
for the charge of the Na$^+$ or Cl$^-$ ions). 
In this work we shall study the three phase equilibria between the methane hydrate (solid), an aqueous solution containing NaCl (liquid) and the methane gas phase using computer simulations. To the best of our knowledge, no one has yet attempted to calculate the complete three-phase coexistence line for the methane hydrate in equilibrium with the aqueous NaCl solution and the gas phase. In 2018, Fernandez-Fernandez {\it et al.}\cite{JML_2019_274_0426} determined a point on the line at seawater conditions and different pressures using a classical unit charge model for ions. Their results with a model of unit charges will help us to compare with those obtained in this work using partial charges.

The work is organized as follows: In section II the simulation details as well as the methodology used are shown. Section III presents the results of the three-phase equilibrium line for the methane hydrate system at different salt concentrations and finally the main conclusions are presented.

\section{\label{sec:meth}Methodology}

Methane hydrate adopts the sI cubic structure. The unit cell is formed by 46 water molecules and 8 methane molecules. In this work we have used a methane hydrate configuration of 3$\times$3$\times$3 unit cells with a total of 1242 water molecules and 216 methane molecules to build the solid slab of the initial configuration. The occupation of all cages is 100\%. The crystallographic positions of methane hydrate were taken from the work of Yousuf {\it et al.}\cite{Appphys-2004-78-925}. Methane hydrates present proton disorder\cite{pauling_book,JCP_1977_66_4699,AC_10_72_1957}. The algorithm proposed by Buch {\it et al.}\cite{JPCB_1998_102_08641} was used in order to generate the initial solid configurations satisfying the Bernal-Fowler rules\cite{JCP_1933_01_00515} and with zero or almost zero dipole moment for the sI hydrate.

We perform direct coexistence simulations at a given pressure and temperature to determine the three-phase coexistence temperature ($T_3$) of a system formed by methane hydrate, NaCl aqueous solution and methane gas.  The direct coexistence method has been implemented due to its robustness and reliability. Previously, this technique have been successfully used to study various ice-systems and hydrate-systems determining coexistence temperatures \cite{JCP_2010_133_064507,JCP_2015_142_124505,PCCP_2017_19_9566,JML_2018_261_513,FPE_2020_513_112548,D1CP02638K,blazquez2022melting,lamas2022freezing,bianco2022phase,JML_2019_274_0426,JCP_2015_142_044501,xiong2020melting,fuentes2014non,jiang2016hydrogen,reddy2016accuracy}. Basically, this technique consists of putting in contact the three phases (solid, liquid and gas). If the temperature of the simulation is above $T_3$ the solid hydrate phase melts and the molality of the aqueous solution decreases (the number of liquid water molecules increases and the number of ions remains unchanged). Otherwise, if the temperature of the simulation is below $T_3$ the hydrate phase grows and the molality of the aqueous solution increases (the number of liquid water molecules is lower and the number of ions is the same). 
Thus, for a given initial salt concentration, $T_3$ is narrow down between the lowest temperature at which the hydrate melts and the highest at which the hydrate phase grows. There is another possible approximation to determine the value of $T_3$, the temperature of the system is fixed and the variation of the molality in the aqueous phase is studied until reaching the equilibrium of the system at that temperature\cite{kim2008effect}. In the present paper we use the two approaches to estimate the complete line of three-phases: narrow down direct coexistence and equilibrium direct coexistence.
We use both approaches to achieve robustness and reliability to our results. Recently, this double approach was successfully performed to determine the phase diagram of the NaCl-water system by Bianco {\it et al.}\cite{bianco2022phase}.

\begin{figure*}[htp]
 \centering
 \includegraphics[width=0.95\textwidth]{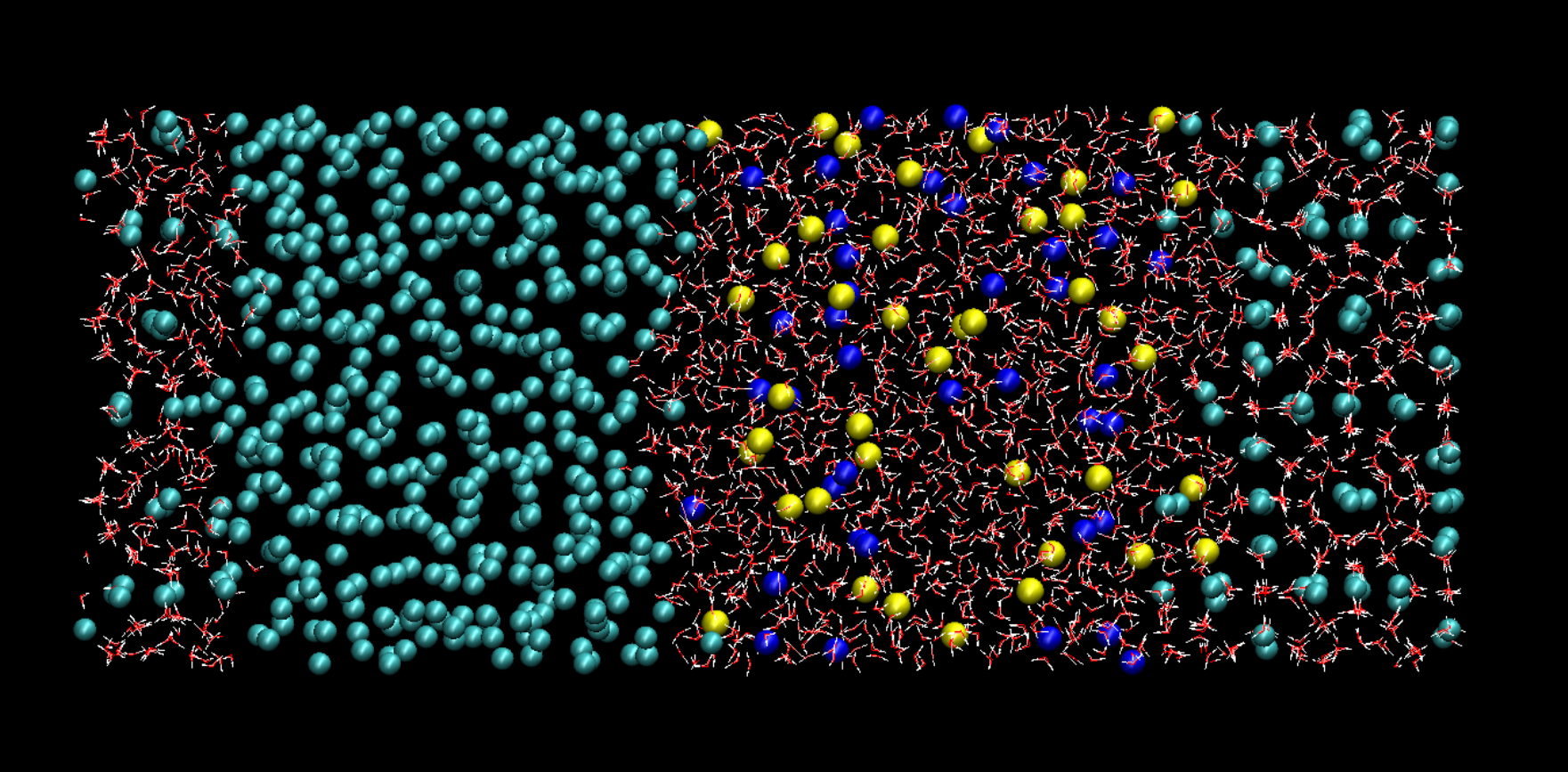}
 \caption{Initial configuration of the three-phase system formed by a slab of methane hydrate in equilibrium with a NaCl solution and a methane gas phase. 
 The aqueous solution of this configuration has a molality of 2 m. 
 Water molecules are represented as sticks in red and white colors, Cl$^-$ and Na$^+$ ions are represented as blue and yellow spheres respectively, and methane molecules are represented as cyan spheres.
 For a clearer visualization of the reading the size of the ions and methane molecules is enlarged with respect to the water molecules. }
  \label{Configinitial}
\end{figure*}

Figure \ref{Configinitial} shows an example of initial configuration used in this work to determine the three-phase coexistence temperature. All the configurations have been generated following the methodology proposed by Fernandez {\it et al.} \cite{JCP_2006_124_144506} and Conde and Vega\cite{JCP_2010_133_064507} to determine equilibrium temperatures by direct coexistence.  
Each initial configuration contains a slab of NaCl aqueous solution surrounded at one side by a slab of solid methane hydrate and at the other side by a slab of methane molecules in gas phase. The initial concentrations of NaCl in water chosen for the present work are: 0.5, 2, 3 and 4 m where m stands for the molality (i.e., mol of salt per kg of water). 
Periodic boundary conditions (PBC) are employed in the three directions of space to ensure that the three phases are in contact with each other. As can be seen in Figure \ref{Configinitial}, the hydrate phase appears at both ends of the image but is continuous through the PBC. The hydrate phase is in equilibrium with the gas phase at one of its interfaces and with the liquid phase at its other interface, resulting in a three-phase equilibrium.

The number of molecules in each phase is detailed in Table \ref{tabla-moleculas}.
For all concentrations studied, the number of water and methane molecules in the hydrate phase remains constant, as do the number of methane molecules in the gas phase and the number of water molecules in the liquid phase. The difference in the number of molecules comes from the number of Cl$^-$ and Na$^+$ ions needed to modify the concentration of the aqueous solution (that is, for molalities of 0.5, 2, 3, and 4 m). Note that in the vicinity of $T_3$, different results may be obtained depending on the initial velocities of the fluid phase molecules. However, this effect can be neglected when using a large number of molecules, as we have done in this work.
In order to study the dependence of the size of the box and the number of molecules of the system with the estimation of the three-phase coexistence temperature, we have replicated by a factor of two in one of the directions of space the liquid phase and the gas phase maintaining the concentration of the NaCl aqueous solution at 2 m. This new configuration is labeled in the Table \ref{tabla-moleculas} with an asterisk. The initial size of the simulation box for each of the initial configurations is 36$\times$36$\times$88 \r{A}$^3$. For the largest system the box size is 36$\times$36$\times$145 \r{A}$^3$. The interfaces between the three phases are perpendicular to the z axis. 

\begin{table}
\caption{\label{tabla-moleculas} 
Number of molecules of the components of the different phases (hydrate, liquid and vapor) in the range of concentrations used in this work. The asterisk ($^*$) corresponds to an initial concentration of 2 m for a larger system studied in this work. }
\begin{ruledtabular}
\begin{tabular}{c c c c c c c c c c c}
Initial molality & &   \multicolumn{3}{c}{Hydrate phase} & &
\multicolumn{3}{c}{Liquid phase} & &
{Vapor phase} \\
     \cline{3-5}
     \cline{7-9}
   mol/kg  & \, & Water & \, & Methane & \, & Water & \, & NaCl  & \,  & Methane \\
\hline
0 & & 1242 & &  211 & & 1110 & & 0 & & 350 \\
0.5 & & 1242 & &  211 & & 1110 & & 10 & & 350 \\
2.0 & & 1242 & &  211 & & 1110 & & 40 & & 350 \\
$^*$2.0 & & 1242 & &  211 & & 2220 & & 80 & & 700 \\
3.0 & & 1242 & &  211 & & 1110 & & 60 & & 350 \\
4.0 & & 1242 & &  211 & & 1110 & & 80 & & 350 \\
\end{tabular}
\end{ruledtabular}
\end{table}

All simulations are performed with GROMACS package of molecular dynamics \cite{spoel05,hess08} in the $NPT$ ensemble and at a fixed pressure of 400 bar. We have employed the leap-frog integrator algorithm\cite{bee:jcp76} with a time step of 2 fs. The temperature is fixed using the Nos\'e-Hoover thermostat\cite{nose84,hoover85} with a coupling constant of 2 ps. 
Anisotropic pressure using the Parrinello-Rahman barostat\cite{parrinello81} with
time constant of 2 ps is applied to the three different sides of the simulation box to allow independent fluctuations and changes in the shape of the solid region, avoiding possible stress in the solid. The cutoff radio employed for van der Waals and electrostatics interactions is 9 \r{A}. Long range energy and pressure corrections to the LJ part of the potential 
 were also applied. The smooth PME method \cite{essmann95} to account for the long-range electrostatic forces is used. The geometry of the water molecules is maintained applying the LINCS algorithm\cite{hess97,hess08b}.

In this work, our system is composed of three components: water, methane and NaCl. Water molecules forming the framework of the solid hydrate and constituting the liquid water phase, gas-phase methane in an independent phase and within the hydrate cages, and NaCl dissolved in the liquid water phase. The choice of a good model for each of the components is crucial to correctly reproduce the experimental results. The description of the pairwise interaction between atoms is given by an electrostatic (Coulombic) contribution and a van der Waals contribution represented by a LJ potential.

 Water molecules in all the simulations carried out in this work have been described using TIP4P/2005 model \cite{abascal05b}. Methane molecules can be described as a single LJ site despite their tetrahedral structure, thus providing similar results to the experimental ones and reducing the computational expense. For these reasons we have selected the parameters proposed by Guillot and Guissuni\cite{JCP_1993_99_8075} and Paschek\cite{JCP_2004_120_6674} to describe methane molecules. However, following the work of Docherty et al.\cite{JPCB_2007_111_8993} we have modified the LJ interactions between methane and water using a deviation from the Lorentz-Berthelot geometric combining rules with a factor $\chi$=1.07 to reproduce the experimental solubility.

The last component of our system is NaCl. In 2019, a force field
for different salts based on the TIP4P/2005 water was proposed by Zeron \textit{et al.}\cite{JCP_2019_151_134504} (denoted as Madrid-2019) and a recent extension for a large variety of salts has been proposed by Blazquez \textit{et al.}\cite{madrid_2019_extended}. These models use the idea of scaling charges of ions to account for the polarization of electrolytes in water. The charge assigned to Na$^+$ and Cl$^-$ is reduced to $\pm$0.85\textit{e}. We have chosen
this model due to its good performance in salt solutions properties such as
the salting out effect of methane\cite{FPE_2020_513_112548}, the freezing depression of ice\cite{lamas2022freezing}, the correct description of the temperature of maximum density\cite{sedano2022maximum} or the improvement of the description of transport properties respect to unit charge models\cite{madrid_2019_extended}.
Furthemore, Joung-Cheatham (JC) model\cite{JoungCheatham,ben:jcp16} with unit charges in combination with TIP4P/2005 water model has also been tested in this work to compare it with the scaled charge model (in particular the model proposed for SPC/E water). 

The parameters used in this work for water, methane and NaCl (using Lorentz-Berthelot combining rules between the ions and CH$_4$) are collected in Table \ref{tabla-modelos}.
Notice that for the Madrid-2019 force field no LB rule is applied for water-ion nor for ion-ion interaction.

\begin{table*}
\caption{\label{tabla-modelos} Force field parameters for water, methane and ions simulated in this work. TIP4P/2005 water parameters are taken 
from Abascal and Vega \cite{abascal05b}. LJ interaction parameters for methane are
taken from Docherty \textit{et al.} \cite{JPCB_2007_111_8993} and Na and Cl ions from Zeron \textit{et al.}\cite{JCP_2019_151_134504} (Madrid-2019 model) and
Benavides \textit{et al.}\cite{ben:jcp16,JoungCheatham} (JC-TIP4P/2005 model). 
The subscript $_{i_{Ow}}$ refers to the cross-interaction of each center with the oxygen of the water molecule. Notice that for the Madrid-2019 force field no LB rule is applied for water-ion nor for ion-ion interaction}
\begin{ruledtabular}
\begin{tabular}{c c c c c c c c c c c}
Model & & Charge & & $\sigma_{ii}$ & & $\epsilon_{ii}$ & & $\sigma_{i_{Ow}}$ & & $\epsilon_{i_{Ow}}$\\
    & & (\textit{e}) & & (\r{A}) & & (kJ/mol) & & (\r{A}) & & (kJ/mol) \\
\hline
TIP4P/2005\\
O  & & 0  & & 3.1589  & &   0.7749 \\
H & & +0.5564  & &  & &  \\
M & & -1.1128  & &  & &  \\
Methane\\
CH$_{4}$  & & 0  & & 3.7300  & & 1.2263 & & 3.4445 & & 1.0430\\
Madrid-2019\\
Na$^+$  & & +0.85 & & 2.21737  & &  1.472356 & & 2.60838 & & 0.793388\\
Cl$^-$  & & -0.85  & & 4.69906  & & 0.076923 & &  4.23867 & & 0.061983  \\
Na$^+$- Cl$^-$  & &   & & 3.00512  & & 1.438894 & &   & &   \\
JC-TIP4P/2005\\
Na$^+$  & & +1.00 & & 2.160  & &  1.47544 & & 2.659 & & 1.06926\\
Cl$^-$  & & -1.00  & & 4.830  & & 0.05349 & &  3.994 & & 0.203603  \\
Na$^+$-Cl$^-$  & &   & & 3.495  & & 0.280935 & &   & &   \\
\end{tabular}
\end{ruledtabular}
\end{table*}

\section{\label{sec:res}RESULTS}

\subsection{\label{subsec:pure}Methane Hydrate in pure water}

In 2010 Conde and Vega \cite{JCP_2010_133_064507} determined the three-phase coexistence temperature ($T_3$) at different pressures for a system composed by methane hydrate in equilibrium with a liquid pure water phase and a methane gas phase for different water models (i.e., TIP4P/2005\cite{abascal05b}, TIP4P/Ice\cite{JCP_2005_122_234511} and TIP4P\cite{tip4p}). At 400 bar, Conde and Vega obtained a value of $T_3$=302(3) K (for a small system) and $T_3$=297(8)K (for a large system) using the TIP4P/Ice model in agreement with the experimental value of 297 K \cite{Sloan_book_hydrates}.
Subsequently, different groups studied this same $T_3$ for methane hydrates using different methodologies or initial conditions. Jensen \textit{et al.}\cite{JPCB_2010_114_5775} (in the same year of publication that Conde and Vega) performed Monte Carlo simulations to determine $T_3$ of methane hydrate by thermodynamic integration obtaining a result approximated to $T_3$=314 K which clearly overestimated the experimental data and those results obtained by Conde and Vega. Michalis \textit{et al.}\cite{JCP_2015_142_044501} also differs from the values of Conde and Vega using the direct coexistence technique reporting a value of 293.4 K. 
Regarding more recent studies, Fernandez-Fernandez \textit{et al.}\cite{JML_2019_274_0426} performed direct coexistence simulations of  methane hydrate in pure water with different system sizes obtaining different values of $T_3$ in function of the number of methane molecules in the gas phase. Their data were closer to those of Michalis \textit{et al.} and underestimated the experimental value. 
Finally, Grabowska \textit{et al.}\cite{grabowska2022solubility} have also recalculated the $T_3$ of the methane hydrate at 400 bar. Nevertheless they have employed two different methods: The first one is the direct coexistence method, obtaining a $T_3$=294(2) K, the second method is based on solubility calculations and provides an estimation of $T_3$=295(2) K.

A compilation of these values at 400 bar is shown in Table \ref{tabla-pure}. As can be seen the more current estimates point out to a value of 294(2) K. 
Notice that the value of T$_3$ depends on the particular choice of the cutoff, and could also have finite size effects and that would explain the differences between the results of different groups (it would be useful in the future to study that in more detail). 
In this work we shall adopt the TIP4P/2005 for water (rather than the more popular in hydrate studies TIP4P/Ice).  The reason for that is that not reliable force field has been proposed for NaCl in water when described with TIP4P/ice.  The Madrid-2019 force field of NaCl was designed for TIP4P/2005 and for this reason we shall use this model in this work. It should be reminded that models with a good melting point of ice I$_h$ predict accurately the value of $T_3$ for hydrates\cite{conde2013note}.
The melting point of ice I$_h$ for TIP4P/Ice and TIP4P/2005 are 270 K and 250 K respectively\cite{conde2017high,blazquez2022melting}. Thus when using TIP4P/2005 we can not expect to reproduce the experimental values of $T_3$. 
However, in this work we focus on the shift in the value of $T_3$ due to the addition of salt rather than in the absolute values of $T_3$. In this work we have recalculated the value of $T_3$ at 400 bar when no salt is present using TIP4P/2005. 

\begin{table}
\caption{\label{tabla-pure} Three-phase coexistence temperature ($T_3$) of methane hydrate obtained for different authors at 400 bar for the TIP4P/Ice and TIP4P/2005 water models. No ions are included in the liquid water phase. The last column corresponds to the year of publication of each reference.  
}
\begin{ruledtabular}
\begin{tabular}{c c c c c c c}
Model & & $T_3$ (K) && Reference && Year \\
\hline
Experimental & &  297 && Sloan\cite{Sloan_book_hydrates}  && 1990 \\
TIP4P/Ice & &  302(3) && Conde et \textit{al.}\cite{JCP_2010_133_064507} && 2010 \\
TIP4P/Ice & &  297(8) && Conde et \textit{al.}\cite{JCP_2010_133_064507} && 2010\\
TIP4P/Ice & &  314(7) && Jensen \textit{et al.}\cite{JPCB_2010_114_5775} && 2010\\
TIP4P/Ice & &  293.4(0.9) && Michalis et \textit{al.}\cite{JCP_2015_142_044501} && 2015 \\
TIP4P/Ice & &  290.5(5) && Fernandez-Fernandez \textit{et al.}\cite{JML_2019_274_0426} && 2019 \\
TIP4P/Ice & &  293.5(5) && Fernandez-Fernandez \textit{et al.}\cite{JML_2019_274_0426} && 2019 \\
TIP4P/Ice & &  294(2) && Grabowska \textit{et al.}\cite{grabowska2022solubility} && 2022 \\
TIP4P/Ice & &  295(2) && Grabowska \textit{et al.}\cite{grabowska2022solubility} && 2022 \\
TIP4P/2005 & &  281(2) && Conde et \textit{al.}\cite{JCP_2010_133_064507} && 2010\\
TIP4P/2005 & &  279(1) && This work && 2022 \\
\end{tabular}
\end{ruledtabular}
\end{table}

\begin{figure}[htp]
 \centering
    \includegraphics[width=0.48\textwidth]{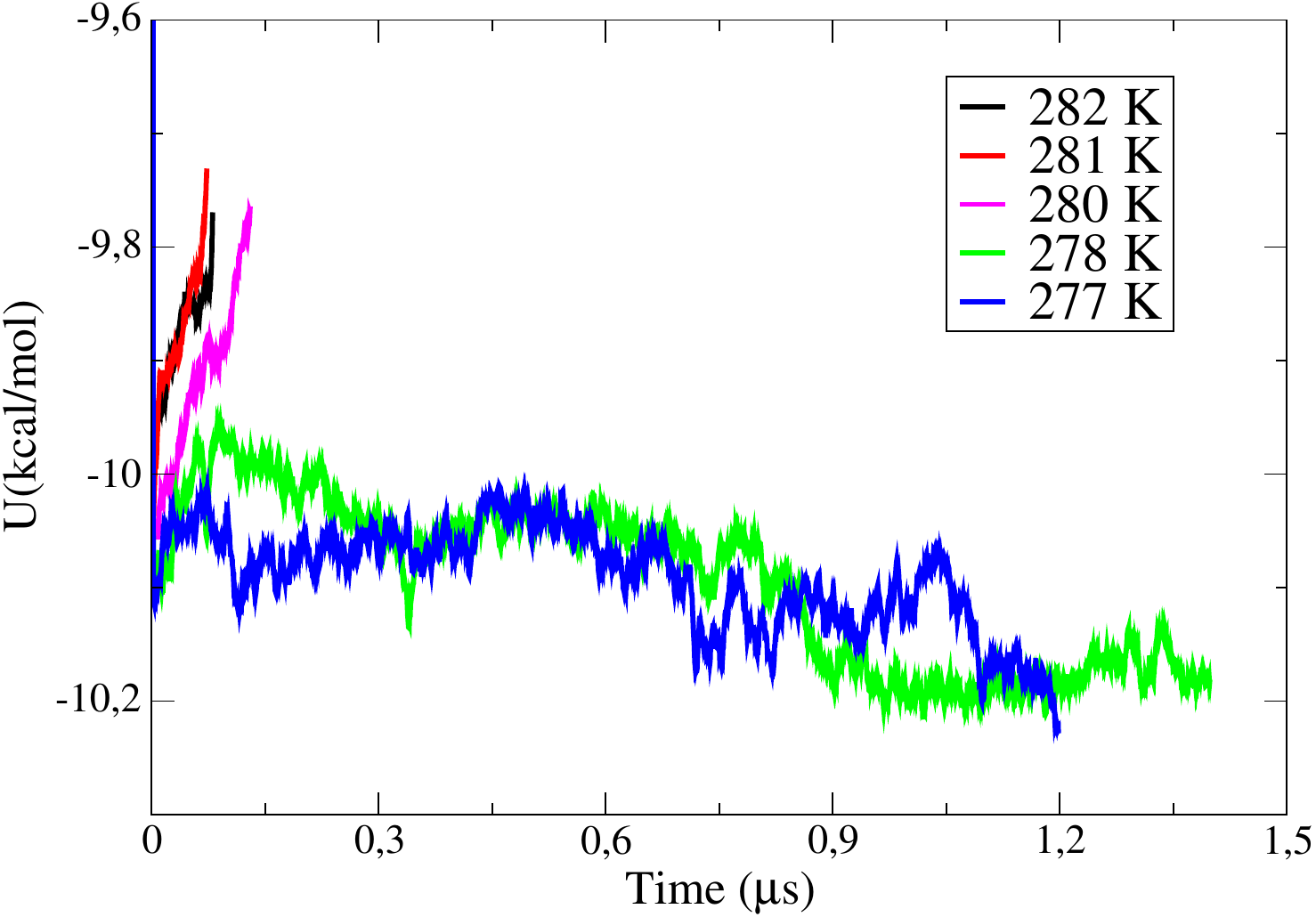}
    \caption{Evolution of the potential energy of the methane hydrate system in pure water described in this work as a function of time for $NpT$ simulations at 400 bar using TIP4P/2005 water potential model. }
    \label{energia-0m}
\end{figure}

To determine $T_3$ of methane hydrate in pure water we follow the procedure described by Conde and Vega\cite{JCP_2010_133_064507} in 2010. It basically consists of putting three coexisting phases in contact and simulating different temperatures for a given pressure. Figure \ref{energia-0m} shows the evolution of potential energy as a function of time. We have used the same initial configuration for all temperatures. It can be seen that the initial energies in the Figure \ref{energia-0m} are not exactly the same in all cases due to quickly relaxation of the system (about 0.1 ns).
Notice that all the potential energies presented in this work are in kcal/mol and per molecule (which means that we have divided the energy by the total number of particles of the system (i.e., water, methane and number of ions).

As it mentioned above, in the $NpT$ simulations each side of the box can fluctuate independently. These non-dependent fluctuations allow the water molecules to accommodate themselves in the solid structure of the hydrate when the system freezes or in the liquid phase when the system melts. At temperatures above $T_3$ (i.e., 280, 281 and 282 K) it can be seen how the potential energy increases with time, indicating the melting of the methane hydrate. Otherwise, if the potential energy decreases with time, the growth of methane hydrate occurs as shown in the Figure \ref{energia-0m} at temperatures 277 and 278 K.

According to these potential energy variations, the lowest temperature at which the hydrate melts is at 280 K and the highest temperature at which the hydrate freezes is at 278 K. With these results about the potential energy we report a value of $T_3$ for the TIP4P/2005 model at 400 bar of 279(1) K. This value is slightly different, although within the error bar, from that obtained more than a decade ago by Conde and Vega\cite{JCP_2010_133_064507} for the TIP4P/2005 model under the same conditions, which was 281(2) K (see Table \ref{tabla-pure}). This new $T_3$ value obtained will be the reference three-phase coexistence temperature used throughout this work when no ions are included in the aqueous phase.

\subsection{\label{subsec:0.5m} Methane Hydrate in 0.5 m aqueous NaCl solution}

Once we have determined the $T_3$ of methane hydrate in pure water we now study the $T_3$ at conditions close to seawater from an initial molality for the NaCl aqueous phase of 0.5 m in the three-phase initial configuration. The selected pressure is 400 bar and we use the Madrid-2019 model to simulate the system of water-NaCl. Note that we chose this pressure to increase the solubility of methane in water, making it possible to observe hydrate growth within a reasonable timeframe that can be studied through simulations lasting microseconds. Furthermore, this pressure allows us to emulate the conditions found in deep oceans.

\begin{figure*}[htp]
 \centering
    \includegraphics[width=0.48\textwidth]{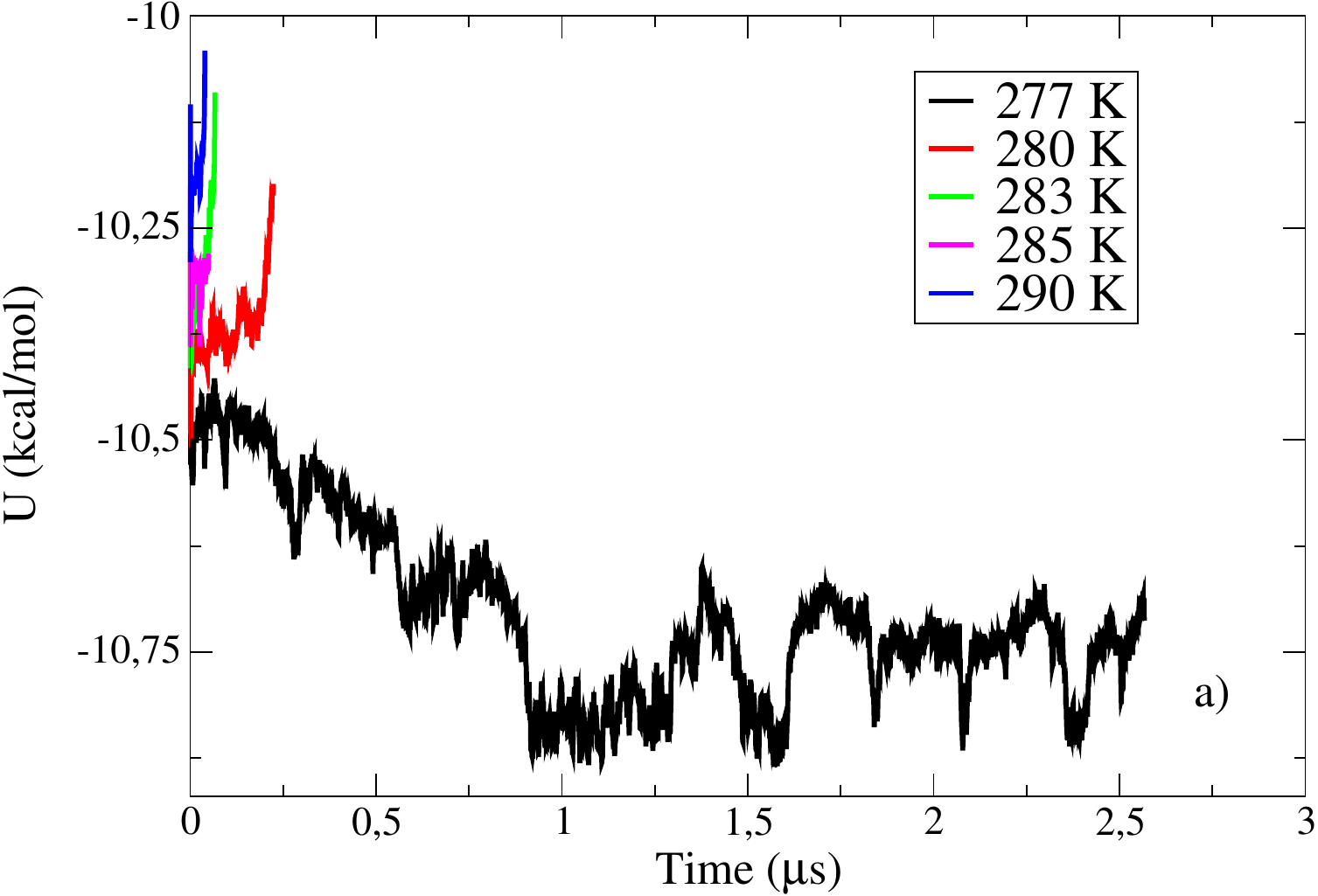}
    \hspace{0.5cm}
    \includegraphics[width=0.45\textwidth]{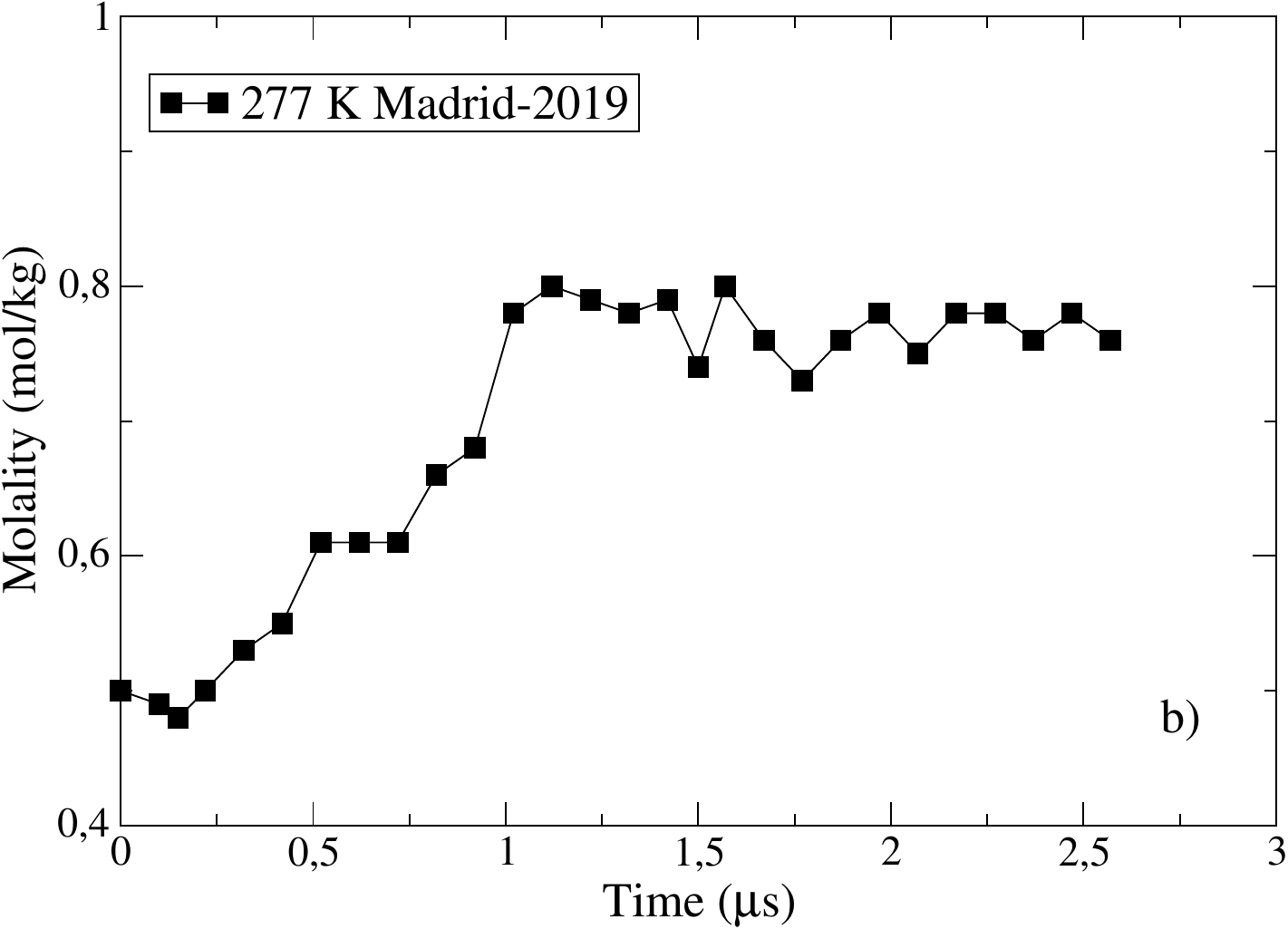}
    \caption{
    a) Evolution of the potential energy as a function of time obtained at several temperatures for the solid methane hydrate-NaCl(aq)-methane gas system with a salt concentration of 0.5 m.
    (narrow down method). 
    b) Evolution of the molality for the NaCl aqueous phase as a function of time calculated at 277 K starting from an initial configuration with a molality of 0.5 m (equilibrium method).
    All simulations were performed using the Madrid-2019 model and at 400 bar.}
    \label{energia-05m}
\end{figure*}

The results obtained at 0.5 m for the evolution of potential energy with time are shown in the Figure \ref{energia-05m}a). Following the same methodology as the one followed for methane hydrate in pure water where $T_3$ is narrowed down, we obtain a value of $T_3$=278.5(1.5) K, only half a degree below the three-phase coexistence temperature in the pure water system. The presence of a small amount of ions in liquid water is practically imperceptible in the freezing depression and is within the error bar. In Figure \ref{energia-05m}a) we observe that $T_3$ is narrowed down between the temperatures of 280 K and 277 K.

\begin{figure}[htp]
 \centering
    \includegraphics[width=0.48\textwidth]{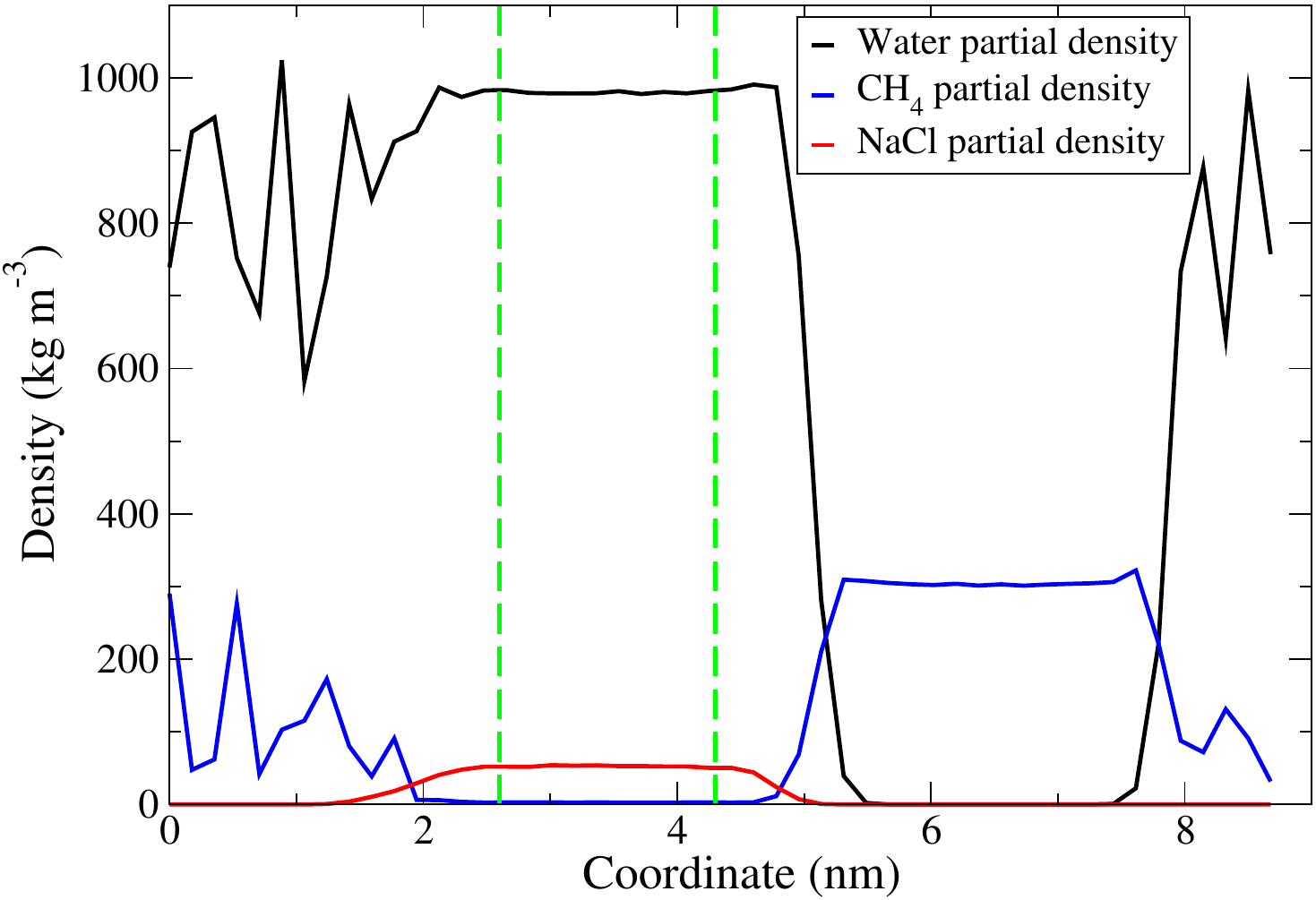}
    \caption{
    Density profile along the direction perpendicular to the interface for the solid methane hydrate-NaCl(aq)-methane gas system at 277 K and calculated after 1 $\mu$s when the system reached the equilibrium.  The interval between the two green dashed lines is used to estimate the molality of the NaCl aqueous phase.}
    \label{partial-density}
\end{figure} 

Until now, we have been determining $T_3$ fixing the initial concentration of salt and performing simulations at different temperatures. However, it is possible to use an alternative methodology. It can be fixed a temperature and performs long simulations until the system reaches equilibrium and estimate the concentration at which the equilibrium is reached. We call this method the equilibrium method to differentiate it from the narrow down method. Previously, this method has been used satisfactorily for NaCl(aq)/ice systems\cite{bianco2022phase,lamas2022freezing}.

It can be seen in the Figure \ref{energia-05m}a) at 277 K how the potential energy of the system decreases until reaching a plateau around 1 $\mu$s. The decrease in potential energy initially reveals the growth of the methane hydrate phase and how equilibrium of the system is subsequently reached. We selected this run at 277 K to study the evolution of the molality of the aqueous phase as a function of time. 

Indeed, the Figure \ref{energia-05m}b) shows how, starting from an initial configuration of 0.5 m, the hydrate phase grows reducing the number of water molecules in the aqueous phase and causing an increase in the molality of the aqueous phase. The molality increases (due to the growth of the hydrate) until the system reaches equilibrium (about 1 $\mu$s). We calculate the molality of the aqueous phase in the equilibrium system as the average value of the molalities from 1 $\mu$s to end of run. Thus, the molality of the system at 277 K and 400 bar is 0.77(02) m. The error bar for the molality calculation was estimated as the standard deviation of the data.

It is important to note that the final molality of the aqueous phase may be different from the initial one as the system approaches equilibrium. The hydrate can grow and increase the apparent concentration (increasing the molality of the aqueous phase) or it can melt and decrease the apparent concentration (decreasing the molality of the aqueous phase).

In Figure \ref{partial-density} we present the density profile for the studied system and how the technique to estimate the molality of the aqueous phase works. 
The three different components of the system (water, methane and NaCl) are shown. The density of each component changes in function of the phase. In the first region of the density profile (up to 2 nm) we observe the solid phase of the hydrate in which water and methane are present (it can be seen also in Figure \ref{Configinitial}). Notice that there is no NaCl in the solid phase, in the case of ices a small amount of the ions can dope the solid phase\cite{PCCP_2017_19_9566,D1CP02638K} but in this case we do not observe this doping process in the hydrate phase. This behavior may be attributed to the fact that the doping process is expected to occur at moderate or deep supercooling conditions (i.e., temperatures lower than those we are simulating). Under high supercooling conditions, the growth rate of the hydrate increases, making the doping process presumably easier to observe due to the more frequent presence of defects. In the central region of the profile we can see the liquid phase of the system which allow us to calculate the concentration of NaCl in the solution. We consider an homogeneous region of the solution (between the vertical green dashed lines) and from the partial densities of water and NaCl we calculate the concentration of NaCl (in mol of NaCl per kg of water). Finally, in the density profile we can also observe the gas phase in which the density of methane is higher and the densities of the other components (i.e., water and NaCl) are almost zero. This procedure is similar for all the configurations studied in this work. Notice that the density profile has been calculated after 1 $\mu$s when the system reaches the equilibrium as we show in Fig. \ref{energia-05m}a) and \ref{energia-05m}b)

\subsection{\label{subsec:2m}Methane Hydrate in 2 m aqueous NaCl solution}

\begin{figure*}[htp]
 \centering
    \includegraphics[width=0.48\textwidth]{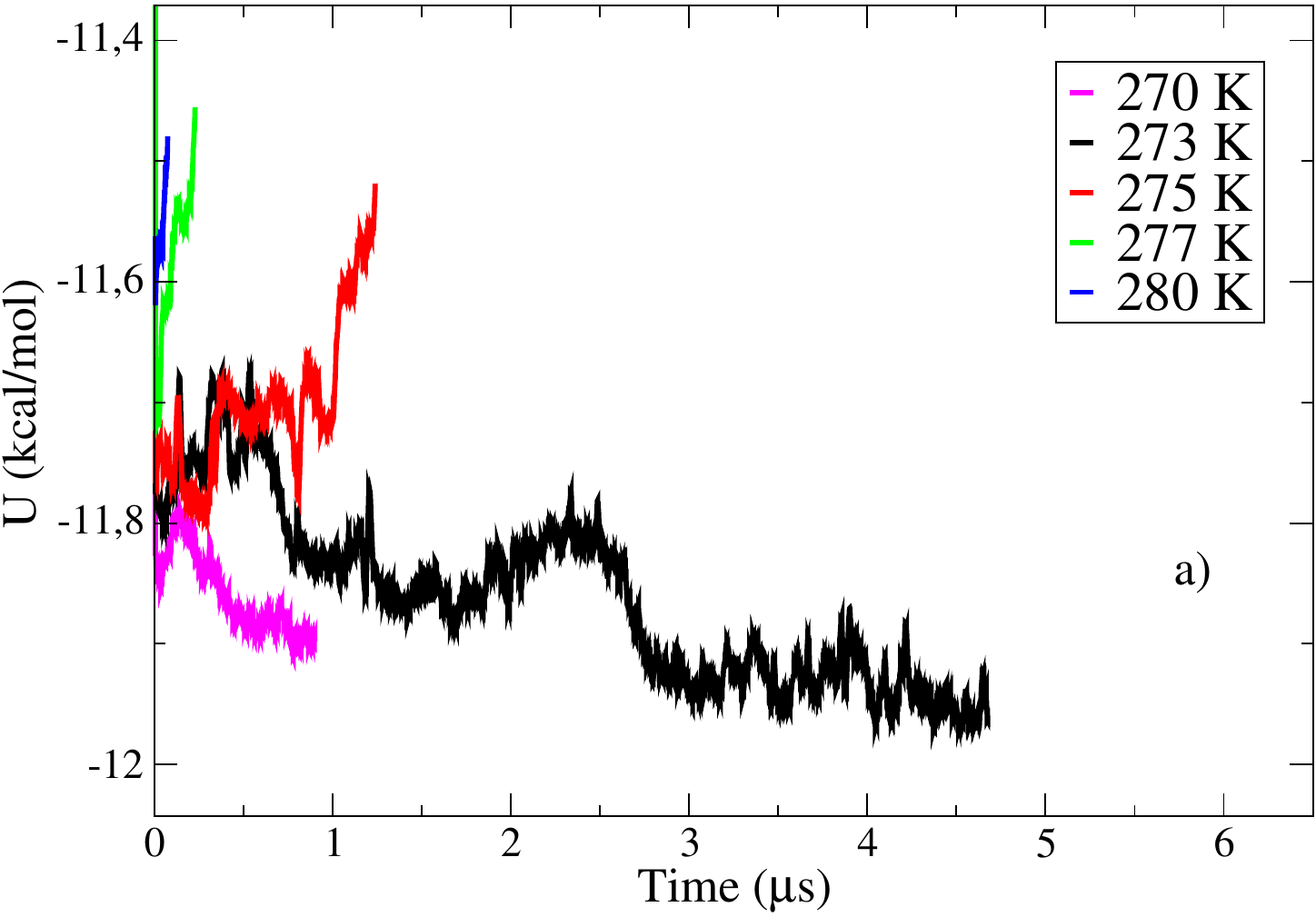}
    \includegraphics[width=0.465\textwidth]{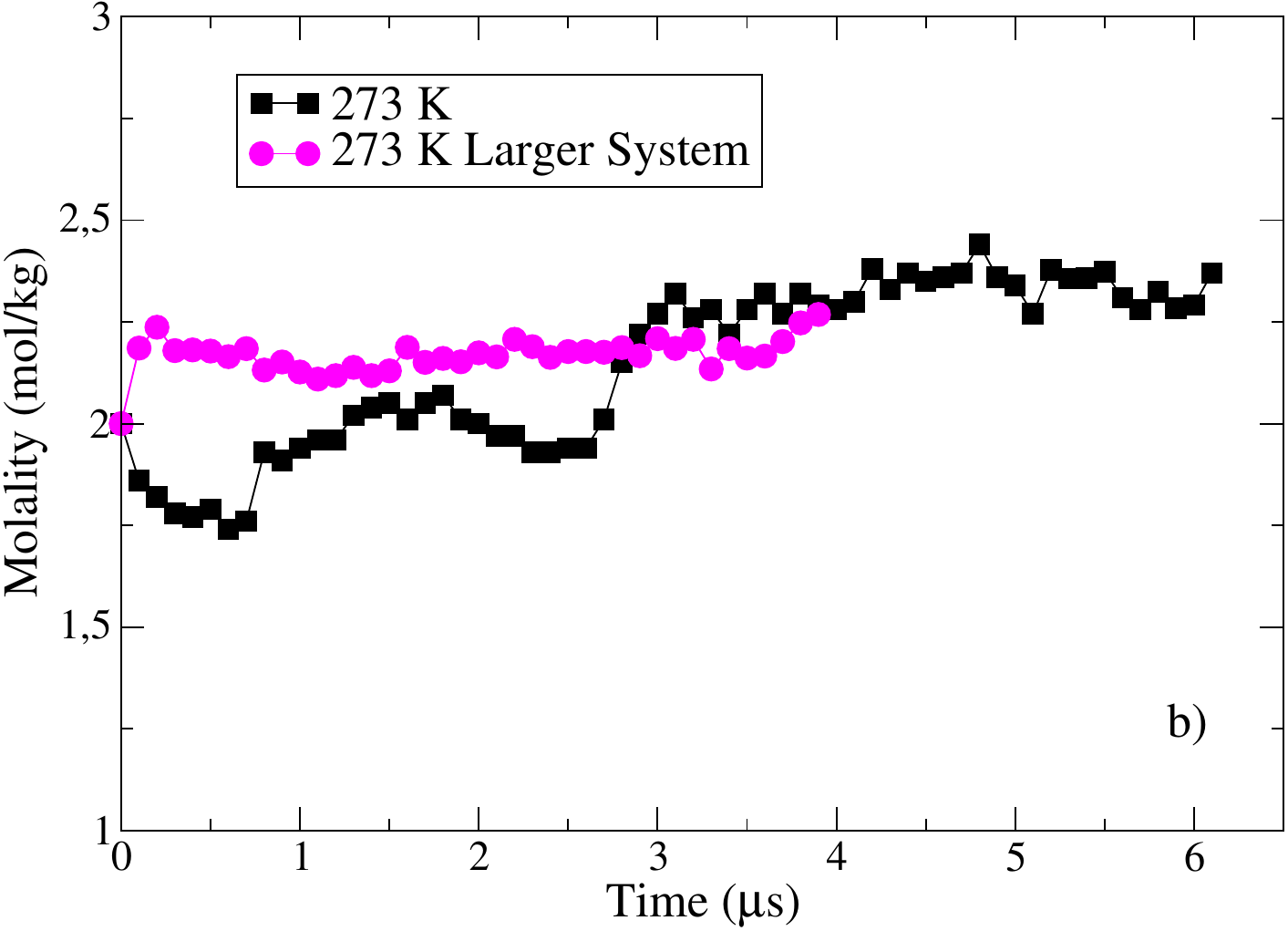}
    \caption{
    a) Evolution of the potential energy as a function of time obtained at several temperatures for the solid methane hydrate-NaCl(aq)-methane gas system with a salt concentration of 2 m.
    (narrow down method). 
    b) Evolution of the molality for the NaCl aqueous phase as a function of time calculated at 273 K starting from an initial configuration with a molality of 2 m (equilibrium method).
    All simulations were performed using the Madrid-2019 model and at 400 bar. At 273 K two sizes of systems were used.}
    \label{energia-2m}
\end{figure*}    

The next salt concentration of the aqueous phase that we study is 2 m. In Figure \ref{energia-2m}a), the potential energy at different temperatures as a function of simulation time for a system formed by a solid phase of methane hydrate in equilibrium with a 2 m NaCl aqueous phase and a methane gas phase is shown. All simulations start from the same initial configuration. For this system, $T_3$ is located between the temperatures of 275 K (lower temperature at which the hydrate phase melts) and 273 K (higher temperature at which the hydrate phase grows). Thus, the value of $T_3$ is 274(1) K for a system with a molality of 2 m for the aqueous phase. This value of $T_3$ is 5 K lower than the system without ions. This amount of ions in the aqueous phase represents a noticeable depression in the $T_3$.

We observe in Figure \ref{energia-2m}a) that the simulation at 273 K reaches a region of stability at long times. For this temperature, the length of the simulation is around 6 $\mu$s  and several abrupt drops in potential energy can be clearly seen, revealing different times in which the hydrate grows. About 700 ns the potential energy decreases and reaches a first plateau. Then, about 2.7 $\mu$s another drop in the energy can be observed indicating a new growth of the hydrate and it remains at a stable plateau until the end of the simulation run. We observe that the system has finally reached equilibrium at a molality of the aqueous phase different from the initial one. We chose this temperature (273 K) to study the evolution of molality with the equilibrium method.

Figure \ref{energia-2m}b) shows the evolution of the molality as a function of time at 273 K starting from an initial configuration where the molality of the NaCl aqueous phase is 2 m. As we had predicted with the potential energy, two remarkable jumps are observed in the system until it reaches equilibrium. In Figure \ref{energia-2m}b) we can associate the two increases in molality with the two decreases in the potential energy. Averaging the values obtained above 3 $\mu$s (from which the system remains in equilibrium for more than 3 $\mu$s) we calculate the molality at 273 K and 400 bar giving a value of 2.31(05) m.
The decrease in energy implies that the hydrate has grown. This in turn causes the amount of water in the liquid phase to decrease and therefore the concentration of NaCl increases (there is the same amount of NaCl but dissolved in less water).

A question arises at this point, what would happen if at the same temperature the size of the system is different? Would the same molality value be obtained? 
Previously, we observed a size dependence in the estimation of the melting point for ice-water systems\cite{conde2017high}. Thus, in this case, we want to clarify whether these effects are present in our system sizes in the methane hydrates study.
In order to answer these questions, we simulate a larger system at the same conditions (i.e., 273 K, 400 bar and an initial configuration with a 2 m NaCl aqueous phase).
The larger system has the same number of molecules in the hydrate phase but the double of molecules in the liquid and the double of molecules in the vapour phase (see Table \ref{tabla-moleculas}). We observe in Figure \ref{energia-2m}b) that both system sizes reach equilibrium at similar values of molality. For the larger system we consider the equilibrium up to 1 $\mu$s and the molality obtained is 2.17(04) m. For the smaller system we calculate the molality from 3 $\mu$s obtaining a value of 2.31(05) m.

The difference between both results is only 6\% percent, which invites us to conclude that there are no considerable size differences, at least for these two selected sizes, not being necessary to use a system with a greater number of molecules in the aqueous phase, and its consequent computational cost, to calculate the molality of the system in equilibrium.

\subsection{\label{subsec:3m}Methane Hydrate in 3 m aqueous NaCl solution}

\begin{figure*}[htp]
 \centering
    \includegraphics[width=0.48\textwidth]{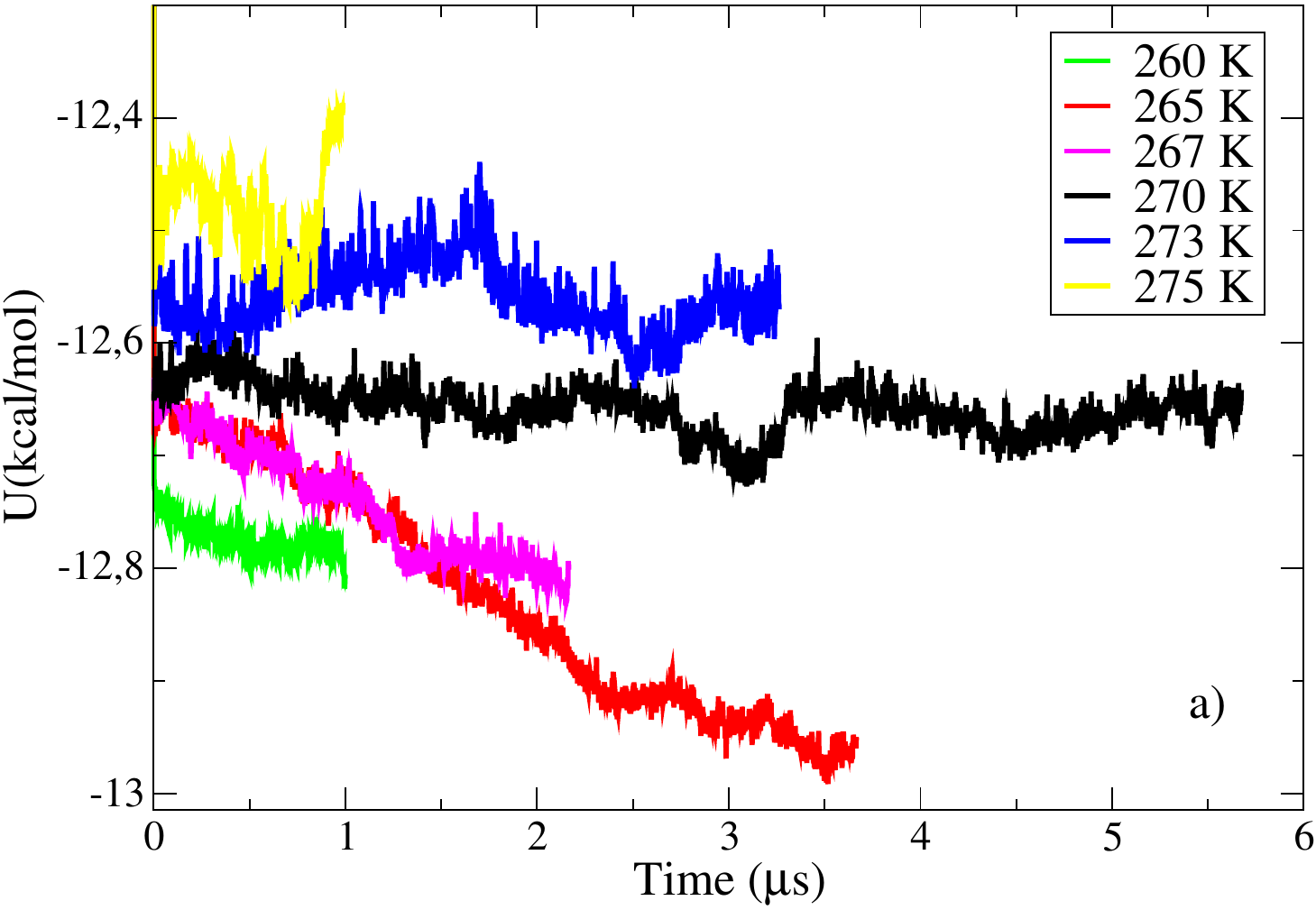}
    \includegraphics[width=0.47\textwidth]{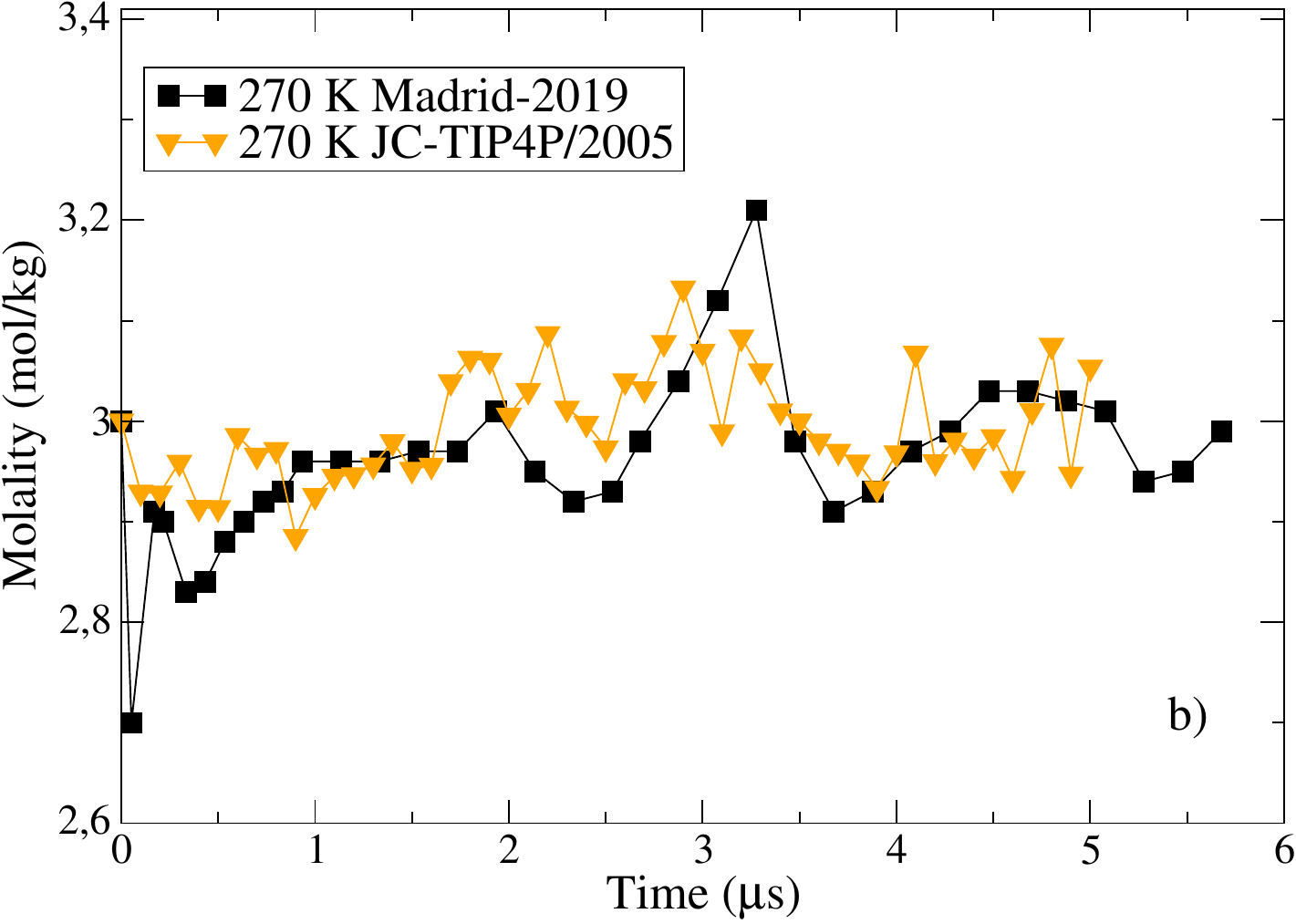}
    \caption{
    a) Evolution of the potential energy as a function of time obtained at several temperatures for the solid methane hydrate-NaCl(aq)-methane gas system with a salt concentration of 3 m (narrow down method). 
    b) Evolution of the molality for the NaCl aqueous phase as a function of time calculated at 270 K starting from an initial configuration with a molality of 3 m (equilibrium method).
    All simulations were performed using the Madrid-2019 model and at 400 bar. We show also the molality results for the unit charges JC-TIP4P/2005 model.}
    \label{energia-3m}
\end{figure*}

The results obtained for the equilibrium of three phases when the molality of the aqueous phase is 3 m are given in Figure \ref{energia-3m}a) and Figure \ref{energia-3m}b). Following the same procedure as in the previous cases we simulate the system at the same pressure and at different temperatures from the same initial configuration. The value of $T_3$ estimated by the narrow down method is 270(3) K. For the equilibrium method, we select 270 K as the temperature at which the molality analysis is calculated since this temperature is maintained with a practically constant potential energy throughout the simulation trajectory (see Figure \ref{energia-3m}a)).

In line with the results obtained for the potential energy, the evolution of molality as a function of time remains practically constant from the beginning of the simulation (see Figure \ref{energia-3m}b)). The chosen temperature of 270 K is a temperature very close to the $T_3$ in these conditions of molality for the aqueous phase of system.

So far, all the simulations in the presence of salt have been carried out with the Madrid-2019 scaled charge model. For this 3 m salt concentration (where the amount of ions present in water is a significant concentration) we compare the molality results obtained by the equilibrium method with the Madrid-2019 scaled charge model and with a unit charge model. The unit charge model chosen is the JC-TIP4P/2005\cite{ben:jcp16} since it is a model that reasonably reproduces the freezing depression line in the ice/NaCl(aq) system.

For both models we start from the same initial configuration and analyze the trajectory obtained at the same temperature (i.e., 270 K) calculating the molality as a function of time. Figure \ref{energia-3m}b) shows the evolution of molality for the Madrid-2019 (scaled charges) and JC-TIP4P/2005 (unit charges). The results obtained are very similar and there are no major differences. However, it is possible that for higher salt concentrations or less complex systems (such as ices) differences between models can be observed\cite{blazquez2023scaled}. In this case the molalities obtained for both models have been calculated starting from 1 $\mu$s. The molality obtained with the Madrid-2019 model is 2.99(07) m and with JC-TIP4P/2005 model is 3.00(05) m.

\subsection{\label{subsec:4m}Methane Hydrate in 4 m aqueous NaCl solution}

\begin{figure*}[htp]
 \centering
    \includegraphics[width=0.48\textwidth]{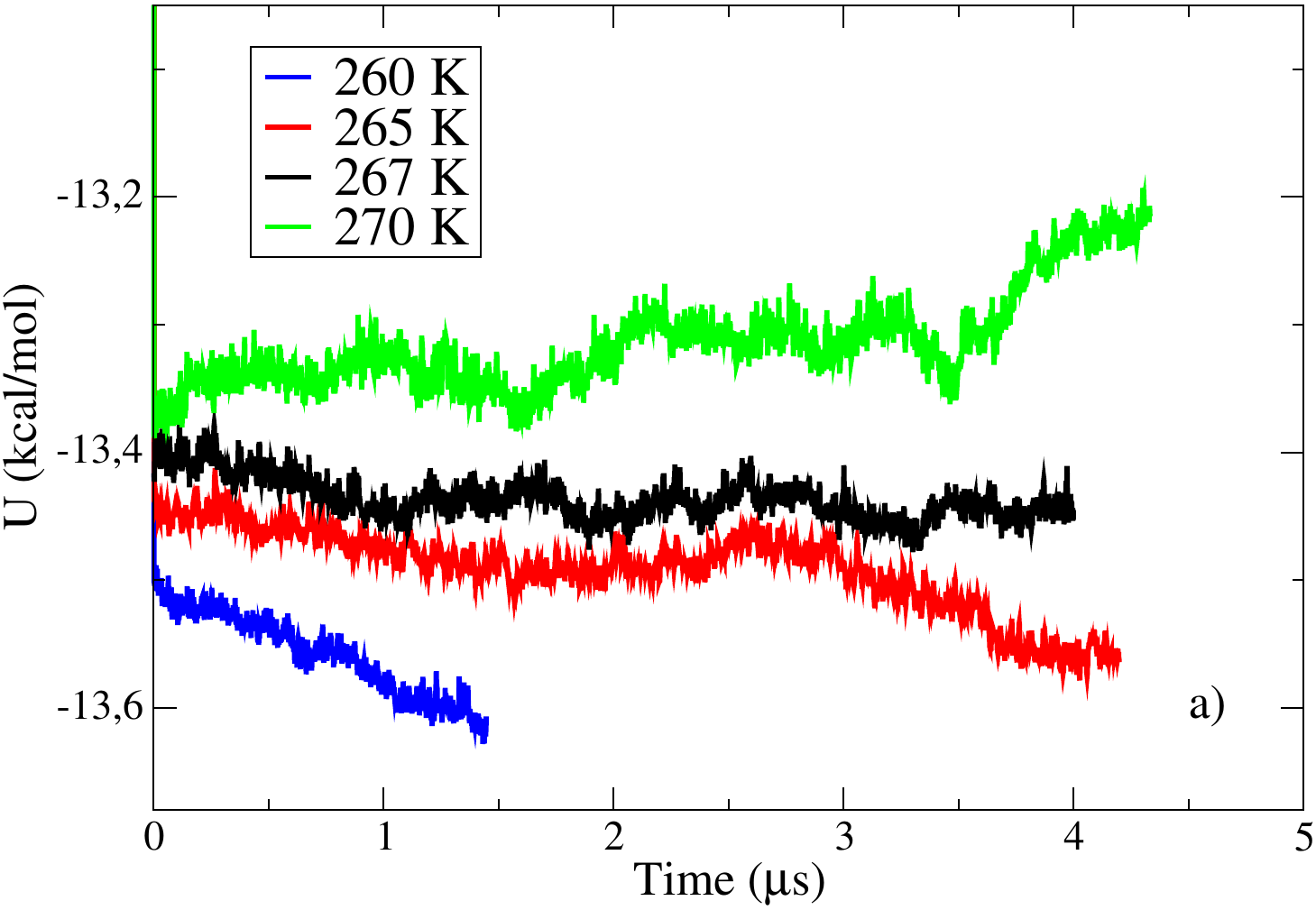}
    \includegraphics[width=0.47\textwidth]{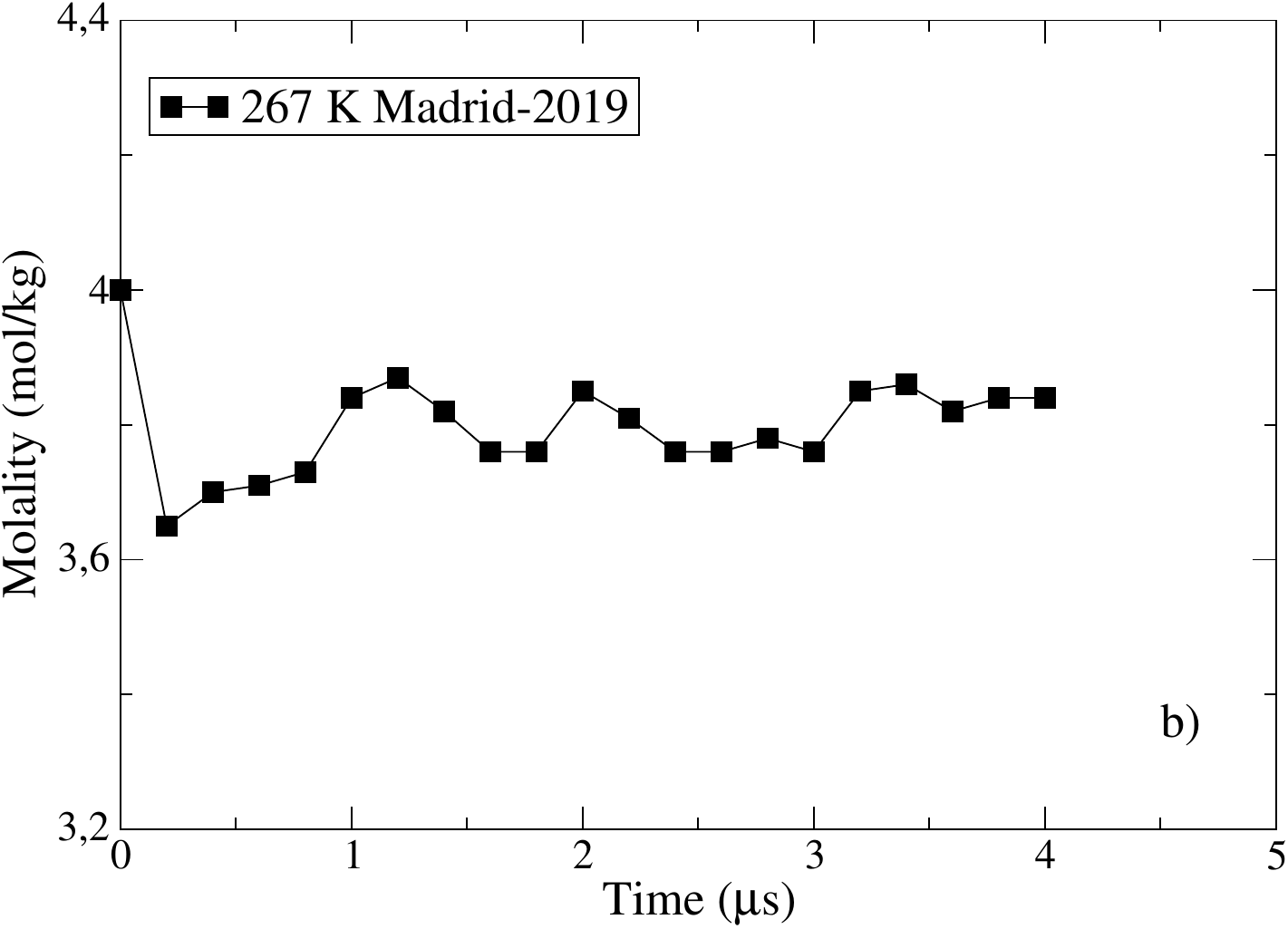}
    \caption{
    a) Evolution of the potential energy as a function of time obtained at several temperatures for the solid methane hydrate-NaCl(aq)-methane gas system with a salt concentration of 4 m (narrow down method). 
    b) Evolution of the molality for the NaCl aqueous phase as a function of time calculated at 267 K starting from an initial configuration with a molality of 4 m (equilibrium method).
    All simulations were performed using the Madrid-2019 model and at 400 bar. }
    \label{energia-4m}
\end{figure*}

Finally, we study the three-phase coexistence with the highest molality, specifically from an initial configuration with a molality of the aqueous phase of 4 m. We perform simulations at the same pressure than in the previous cases and several temperatures. The results of the evolution of the potential energy as a function of time are shown in Figure \ref{energia-4m}a). Clearly at 260 and 265 K it is seen that there is a drop in energy indicating the slow growth of the methane hydrate phase. Moreover, at 270 K the energy rises caused by the partial melting of the hydrate phase. At 267 K the potential energy slightly decreases in the first microsecond of the simulation, remaining constant until 4 $\mu$s. Note that in this case, to obtain simulations of more than 4 $\mu$s, the calculation time was extremely high due to the slow kinetics of the system when the salt concentration is so high. In view of the results in Figure \ref{energia-4m}a), the value of $T_3$ for the 4 m system is 267(3) K. 

In addition, we represent the evolution of molality as a function of time in Figure \ref{energia-4m}b). For this system and after examining different temperatures, we choose 267 K since it seems to be the closest temperature to the $T_3$ as we show in Figure \ref{energia-4m}a). Starting from an initial configuration with a molality of 4 m, the system at 267 K clearly decreases its molality up to 1 microsecond when it slightly increases its molality again and remains constant until the end of the simulation. Averaging the molality from 1 $\mu$s, we obtain a value of 
m=3.81(04) mol/kg for this system with the Madrid-2019 model at 267 K and 400 bar.

For this temperature, the molality of the system under equilibrium conditions (3.81 m) is smaller than the molality of the initial configuration (4 m). This means that part of the initial block of methane hydrate melts, thus increasing the number of water molecules in the aqueous phase and therefore decreasing the molality of the aqueous solution. 

Nevertheless, the chosen temperature of 267 K is  close to the equilibrium of three phases and the molality is very close to the value of the initial configuration. Likewise, both methods, the narrow down method and the equilibrium method are valid and reliable to determine the $T_3$ of all the concentrations studied in this work.

\subsection{T$_3$ equilibrium line for the methane hydrate-NaCl(aq)-
methane gas system}

A summary of all the values of $T_3$ obtained at 400 bar for both the narrow down method and the equilibrium method are shown in Tables \ref{tabla-temperaturas-acot} and \ref{tabla-temperaturas-eq} respectively. The range of salt concentrations studied in the aqueous phase goes from 0.5 to 4 m. 

We also present $\Delta$$T_3$ as the depression of the three-phase coexistence temperature of the system when there are dissolved ions in the aqueous phase ($\Delta$$T_3$ is the difference between $T_3$ in salt solution and $T_3$ in pure water). 
We use this magnitude ($\Delta$$T_3$) to compare with experimental data because of the water model used (TIP4P/2005) does not provide a value of the melting temperature of ice ($T_m$) close to the experimental value as we have mentioned previously. This value is shifted by about 20 K below the experimental value. Certainly other models such as TIP4P/Ice predict the experimental value of $T_m$ (and therefore $T_3$) since it has been fitted to describe this specific property. However, to the best of our knowledge, there are no salt force fields adapted to this model.

In Table \ref{tabla-temperaturas-acot} the concentration of the aqueous phase is fixed and different temperatures are studied estimating the value of $T_3$ by the narrow down method. In the equilibrium method (see Table \ref{tabla-temperaturas-eq}), by contrast, for an initial configuration with an initial molality of the aqueous phase, the temperature is fixed and the molality of the system ($m_{eq}$) is determined when it reaches equilibrium.

\begin{figure}[htp]
    \centering
    \includegraphics[width=0.48\textwidth]{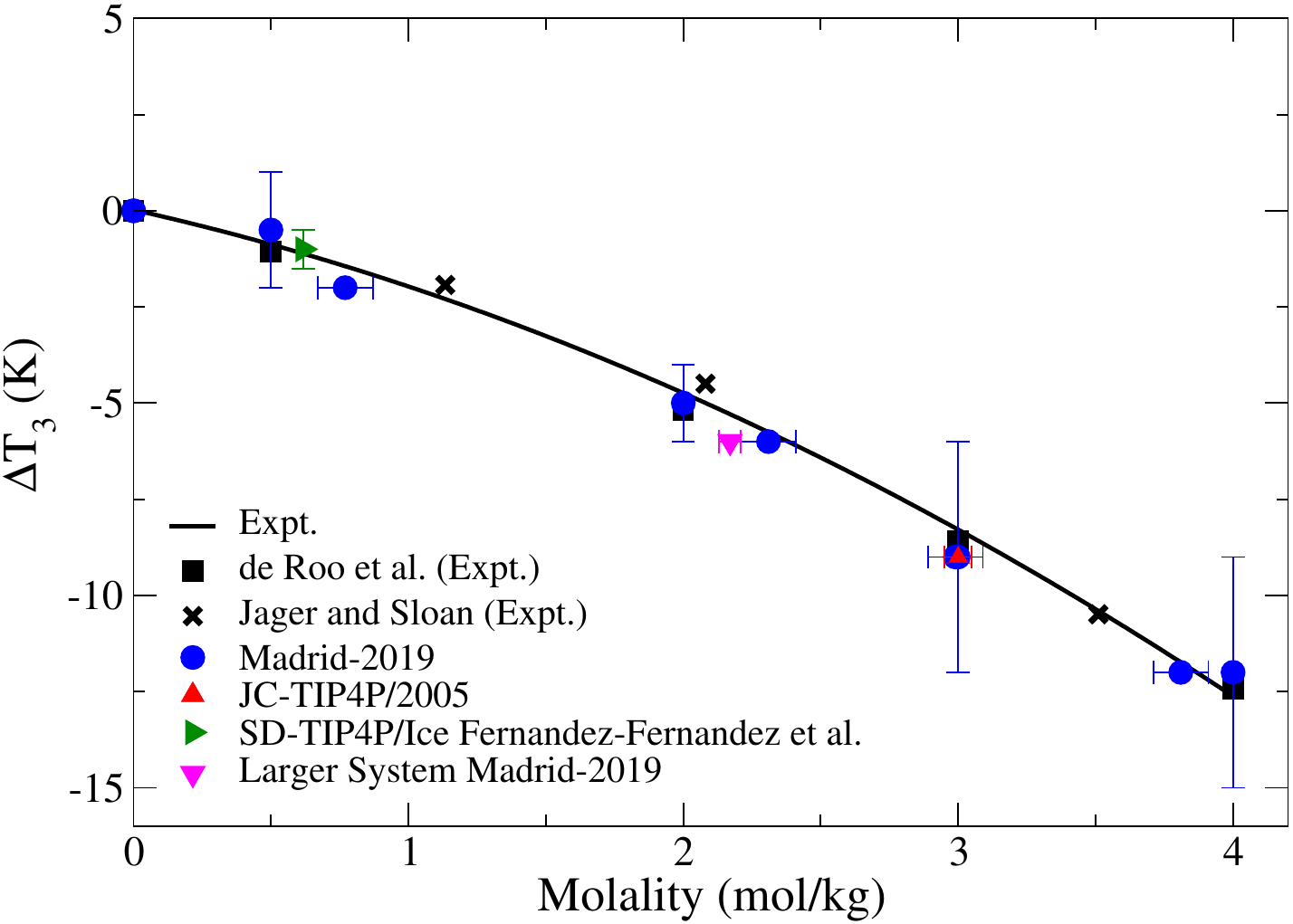}
    \caption{
    Three-phase coexistence line and coexistence point depression for the system formed by a slab of methane hydrate, a NaCl aqueous solution and a methane gas phase as a function of molality of the aqueous phase at 400 bar. 
    $\Delta$$T_3$ is given relative to $T_3$ of methane hydrate in pure water calculated in this work (279 K) for the TIP4P/2005 model.
    Blue circles: results for the Madrid-2019 model. 
    Pink triangle down: results for the Madrid-2019 model using a larger system. 
    Red triangle up: results for the JC-TIP4P/2005 model. 
    Green triangle right: simulation result taken from Fernandez-Fernandez {\it et al.}\cite{JML_2019_274_0426}.
    Solid black line: Fitted experimental data obtained from Roo \textit{et al.}\cite{Aiche_1983_29_651} (black squares) and Jager and Sloan\cite{JAGER200189} (black crosses).}
    \label{deltat}
\end{figure}

And lastly, we conclude with the representation of the three-phase equilibrium line as a function of the salt concentration for the system formed by a methane hydrate phase, an NaCl aqueous phase and a methane gas phase. The equilibrium line is collected in Figure \ref{deltat}. We represent $\Delta$$T_3$ versus molality of the aqueous phase.
As it can be seen in Figure \ref{deltat} when the salt concentration increases, $\Delta$$T_3$ decreases. This behavior also occurs in ice (phenomenon known as freezing depression). In a hydrate-type system, as in ice, the presence of ions in the aqueous phase causes a depression of the three-phase coexistence line.

For the entire range of salt concentration studied, the Madrid-2019 model correctly predicts the equilibrium line in agreement with the experimental fit based on the experimental data of Roo \textit{et al.}\cite{Aiche_1983_29_651} and Jager and Sloan\cite{JAGER200189}. Likewise, there is consistency in the simulation results obtained by the two methods used in this work (narrow down and equilibrium), providing robustness and reliability to our results. It is true and it can be seen in  Figure \ref{deltat} that at high salt concentrations the error bar estimated by the narrow down method for the value of $\Delta$$T_3$ is large due to the increasingly slow kinetics of the system. However, these values at high molalities (3 and 4 m) are in agreement with the results obtained by the equilibrium method and fall within the experimental equilibrium line. Thus, although the error is high by the narrow down method (especially at high molalities), we are certain that the values are trusted.

At molalities close to marine salinity, we have included in the equilibrium line the data reported by Fernandez-Fernandez \textit{et al.}\cite{JCP_2006_124_144506} They use the Smith and Dang unit charge model \cite{SmithDang} for ions in combination with the TIP4P/Ice water model \cite{JCP_2005_122_234511} giving very good result at this salinity concentration. However, in previous work on freezing depression of ice \cite{JML_2018_261_513,bianco2022phase}, it was observed both for unit charge models and for scaled charge models that at low molalities it is very difficult to find differences between force fields because the drop in temperature at those molalities is very small (below of 2 K) being very close to the error bar of the simulation technique itself. Furthermore, it would be interesting to study force fields for sodium chloride in combination with the TIP4P/Ice model along the equilibrium line since this water model reproduces well the three-phase coexistence temperature for methane hydrate. To date there is no suitable combination for this water model.

Additionally, as it can be seen in Figure \ref{deltat} and in Table \ref{tabla-temperaturas-eq}, two sizes of systems studied starting from an initial molality of 2 m provide similar results. Although it is true that more work in this sense should be carried out in order to extract a clear and definitive message about the role of finite size effects in the direct coexistence simulations for the determination of coexistence temperatures.

And finally, the result obtained with the JC-TIP4P/2005 unit charge model is compared with the Madrid-2019 scaled charge model by the equilibrium method. Both values for the temperature of 270 K are practically identical and slightly fall below the experimental line (see Table \ref{tabla-temperaturas-eq} and Figure \ref{deltat}). Nonetheless, at 4 m the scaled charge model correctly predicts the point of coexistence on the experimental line. More work would be needed at high molalities for unit charge models to know if this slight underestimation persists or if it is maintained as it occurs for less complex systems such as ice coexisting with a NaCl solution\cite{JML_2018_261_513,bianco2022phase}.

\begin{table}
\caption{ Three-phase coexistence temperature ($T_3$) and coexistence point depression ($\Delta$$T_3$) for the methane hydrate/NaCl(aq)/methane system studied for different salt concentrations obtained by the narrow down method using the Madrid-2019 model and 400 bar. The concentration of NaCl in the aqueous phase is shown in molality ($m$). 
$\Delta$T$_3$ is the depression of the three-phase coexistence temperature of the system when there are dissolved ions in the aqueous phase. 
The value of $T_3$ for the methane hydrate system in pure water is calculated for the TIP4P/2005 model. The unit of the molality is given in mol/kg and the unit of the temperature is given in Kelvin (K). 
The results obtained in this work are compared with those obtained from a fit of the experimental values of Refs. \cite{Aiche_1983_29_651,JAGER200189}.}
\label{tabla-temperaturas-acot}
\begin{ruledtabular}
    \begin{tabular}{c c c c c c }
Model &  $m$ &  \multicolumn{2}{c}{Simulation} & 
\multicolumn{2}{c}{Experimental} \\
     \cline{3-4}
     \cline{5-6}
 &    &  $T_3$  & $\Delta$$T_3$   & $T_3$ & $\Delta$$T_3$  \\
\hline
TIP4P/2005  & 0 &  279 &   0 & 300.77 & 0  \\
Madrid-2019 &  0.5  &   278.5(1.5) &  -0.5  & 299.70 & -1.067  \\
Madrid-2019  & 2.0  &  274(1)  & -5 & 295.58 & -5.19  \\
Madrid-2019  & 3.0 &  270(3)  & -9  & 292.17 & -8.59 \\
Madrid-2019  & 4.0  &  267(3)  & -12  & 288.34 & -12.43  \\
    \end{tabular}
\end{ruledtabular}
\end{table}


\begin{table}
\caption{\label{tabla-temperaturas-eq} 
Equilibrium molality ($m_{eq}$), three-phase coexistence temperature ($T_3$) and coexistence point depression ($\Delta$$T_3$) for the methane hydrate/NaCl(aq)/methane system obtained by the equilibrium method using the Madrid-2019 model and 400 bar. The initial molalities of each system are 0, 0.5, 2, 3 and 4 mol/kg.
$\Delta$T$_3$ is the depression of the three-phase coexistence temperature of the system when there are dissolved ions in the aqueous phase. 
The value of $T_3$ for the methane hydrate system in pure water is calculated for the TIP4P/2005 model. 
The asterisk ($^*$) represents the result obtained for a larger system and an initial molality of 2 m.
For the system with an initial molality of 3 m, the results obtained for the Madrid-2019 and Joung-Cheatham models are compared. 
The unit of the molality is given in mol/kg and the unit of the temperature is given in Kelvin (K). The results obtained in this work are compared with those obtained from a fit of the experimental values of Refs. \cite{Aiche_1983_29_651,JAGER200189}.}
\begin{ruledtabular}
\begin{tabular}{c c c c c c  }
Model  & $m_{eq}$ & \multicolumn{2}{c}{Simulation} & 
\multicolumn{2}{c}{Experimental} \\
     \cline{3-4}
     \cline{5-6}
 &      & $T_3$ & $\Delta$$T_3$  & $T_3$ & $\Delta$$T_3$ \\
\hline
TIP4P/2005    & 0        & 279 &  0  & 300.77 & 0  \\
Madrid-2019   & 0.77(05) & 277 & -2 & 299.27 & -1.5 \\
Madrid-2019   & 2.31(05) & 273 & -6 & 294.97 & -5.8 \\
Madrid-2019$^*$  & 2.17(04) & 273 & -6 & 295. 47 & -5.3 \\
Madrid-2019   & 2.99(07) & 270 & -9 & 292.47 & -8.3 \\
JC-TIP4P/2005 & 3.00(05) & 270 & -9 & 292.47 & -8.3 \\ 
Madrid-2019   & 3.81(04) & 267 & -12 & 289.07 & -11.7\\
\end{tabular}
\end{ruledtabular}
\end{table}

\section{\label{sec:concl}CONCLUSIONS}

In this work we have estimated the three-phase coexistence temperature for a
methane hydrate system in equilibrium with a NaCl solution and a methane gas phase by using molecular dynamics simulations. This is the first time (to the best of our knowledge) that this three-phase equilibrium line is calculated by computer simulation and the results are in good agreement with the experimental measurements.

In order to estimate the three-phase temperatures ($T_3$) we have employed two different methods (i.e., the equilibrium method and the narrow down method). In the first one, we study the evolution of the concentration of the aqueous phase with time until the system reaches equilibrium by fixing the temperature and pressure. Therefore, we only need a long simulation (on the scale of microseconds) at the fixed thermodynamic conditions. The second method is based on simulating different temperatures for a fixed pressure where the evolution of potential energy as a function of time is monitored, delimiting the equilibrium temperature as that located between the lowest temperature that increases energy and the highest temperature that decreases the potential energy of the system. By studying the evolution of potential energy as a function of time it is possible to narrow down the three-phase equilibrium temperature.

Both methods obtain similar results and allow the prediction of the entire range of concentrations of the aqueous phase. The equilibrium method is faster and with lower uncertainty (although the simulation is longer, there is only one simulation to perform and reach equilibrium of the system). In the narrow down method, the error bars at higher concentrations are larger than in the equilibrium method because at high concentrations it becomes increasingly more expensive to approach equilibrium temperature due to the slow dynamics of the system. However, when the equilibrium region of the system is not known, the narrow down method is very useful to estimate a first starting point on the equilibrium line that serves as a guide.

We have also studied two different force fields for salts (i.e., the Madrid-2019 with scaled charges and the JC-TIP4P/2005 with unit charges) and we have not found differences between these models, obtaining practically the same value. It is possible that for higher salt concentrations differences between models can be observed but the simulations at higher molalities are computationally expensive and the main purpose of this work is not to examine differences between force fields but determine the three-phase line of methane hydrate in the presence of salt.

Moreover, we have studied the role of finite size effects in the estimation of $T_3$ simulating two systems with different numbers of molecules under the same conditions of pressure, temperature and molality. We have observed that there is a slight decrease in the equilibrium molality (at the same temperature) when we use the larger system. However, further efforts are needed to analyze the finite size effects on $T_3$ in hydrate systems.

In view of the results and with the current computational power we conclude that it is possible to study by simulation the $T_3$ of methane hydrate in aqueous systems with salt as shown in this work. Likewise, the Madrid-2019 model gives a very good description of the shift in the equilibrium line. Similarly, the unit charge force field also provides a good estimate of $T_3$ (at least up to molalities above 3 m where the difference between unit charge and scale charge force fields is noticeable).

\begin{acknowledgments}
We would like to congratulate Prof. Holovko for his 80th anniversay and for his many influential contributions in the statistical mechanics of electrolytes. 
This work was funded by Grant No. PID2019-105898GA-C21 and PID2019-105898GA-C22 of the MICINN and by Project No. ETSII-UPM20-PU01 from ``Ayudas Primeros Proyectos de la ETSII-UPM''. M.M.C. acknowledges CAM and UPM for financial support of this work through the CavItieS project No. APOYO-JOVENES-01HQ1S-129-B5E4MM from ``Accion financiada por la Comunidad de Madrid en el marco del Convenio Plurianual con la Universidad Politecnica de Madrid en la linea de actuacion estimulo a la investigacion de jovenes doctores'' and CAM under the Multiannual Agreement with UPM in the line Excellence Programme for University Professors, in the context of the V PRICIT (Regional Programme of Research and Technological Innovation). The authors gratefully acknowledge the Universidad Politecnica de Madrid (www.upm.es) for providing computing resources on Magerit Supercomputer.
\end{acknowledgments}

\providecommand{\noopsort}[1]{}\providecommand{\singleletter}[1]{#1}%

\end{document}